# An Image Encryption Scheme Based on Chaotic Logarithmic Map and Key Generation using Deep CNN

Uğur Erkan[a*], Abdurrahim Toktas[b], Serdar Enginoğlu[c], Enver Karabacak[d], Dang N. H. Thanh[e]

[a] Department of Computer Engineering, Engineering Faculty, Karamanoğlu Mehmetbey University, 70200, Karaman, Turkey

[b] Department of Electrical-Electronic Engineering, Faculty of Engineering, Karamanoğlu Mehmetbey University, 70200, Karaman, Turkey

[c] Serdar Enginoğlu, Department of Mathematics, Faculty of Arts and Sciences, Çanakkale Onsekiz Mart University, Çanakkale, Turkey

[d] Engineering Faculty, Department of Electrical and Electronics Engineering, Marmara University, 34722, Göztepe, Istanbul, Turkey

[e] Department of Information Technology, School of Business Information Technology, University of Economics Ho Chi Minh City, Ho Chi Minh City, Vietnam

[*]ugurerkan@kmu.edu.tr

**Abstract:** A secure and reliable image encryption scheme is presented in this study. The encryption scheme hereby introduces a novel chaotic log-map, deep convolution neural network (CNN) model for key generation, and bit reversion operation for the manipulation process. Thanks to the sensitive key generation, initial values and control parameters are produced for the hyperchaotic log-map, and thus a diverse chaotic sequence is achieved for encrypting operations. The scheme then encrypts the images by scrambling and manipulating the pixels of images through four operations: permutation, DNA encoding, diffusion, and bit reversion. The encryption scheme is precisely examined for the well-known images in terms of various analyses such as keyspace, key sensitivity, information entropy, histogram, correlation, differential attack, noisy attack, and cropping attack. To corroborate the scheme, the visual and numerical results are even compared with available outcomes of the state of the art. Therefore, the proposed log-map based image encryption scheme is successfully verified and validated by the superior absolute and comparative results.

## 1. Introduction

In line with progress in computer and network technologies, real-time messaging and data transferring have become inevitable, particularly for medical, military, and education applications [1–3]. In general, the data is transferred through a wide area network (WAN), which is likely exposed to cyber threats such as network attacks, denial-of-service, man-in-the-middle, and phishing [4]. Therefore, it should be transferred after encrypting the data for providing information security, making the data encryption techniques the most crucial task [5,6]. The well-known encryption techniques that especially emerged for text encryption are data encryption standard (DES), triple-DES (3DES), international data encryption algorithm (IDEA), and an advanced encryption standard (AES). In case they are implemented to the image encryption, the security is low due to high correlation and a large amount of data [7,8].

Image encryption techniques based on the spatial domains such as Deoxyribonucleic Acid (DNA) coding, chaos, cellular automata, and compressed sensing and the frequency domains like Fourier and wavelet transform have been widely exploited for the last decades [9,10]. Chaos-based schemes are the most employed techniques due to high randomization, complexity, sensitivity to initial conditions, and system parameters. Chaos-based image encryption schemes are processed through two stages, in general. In the first stage, the main key is constituted with XOR operation between a secret key and a public key. The image is then encrypted in the second stage by using the main key. It is critical to use an encryption scheme sensitive to the main key to achieve high secure and reliable ciphertext image. Therefore, a key derived from the plaintext image is a quite reasonable and effective way. In the second stage, the image's pixels are interchanged in position and manipulated in color tonal intensity using permutation and diffusion operations, respectively. The permutation and diffusion operations are carried out using chaotic maps with various dimensions and DNA coding managed by the main key. The operations are conducted by a sequence produced from the chaotic map with the initial value and control parameter. Therefore, chaotic maps play a decisive role in achieving success in an encryption scheme. The performance of encryption schemes is evaluated using different analysis approaches such as keyspace, key sensitivity, information entropy, histogram, correlation, differential attack, noisy attack, and cropping attack [11].

Recently, several image encryption schemes have been proposed in the literature [12,13,22–27,14–21]. In [12], an image encryption scheme has been suggested a 2D Logistic-adjusted-Sine map for permutation and introduced a 1D chaotic map for diffusion. In [13], the authors have propounded a scheme based on the integer-based key generation with a modified logistic map. In [14], a 1D Sine powered chaotic map has been developed for image encryption. In [15], a polynomial 1D chaotic map was constituted by combining several chaotic maps. In [16], the authors have introduced a cosine-transform-based chaotic map. In [17], a chaotic map has been obtained by integrating three basic 1D maps conducting cascade, nonlinear combination, and switch operations. [18] has proposed a method based on YoloV3 object detection and chaotic image map. In [19], the authors have defined a scheme combining two Sine maps in the permutation stage and utilizing key-streams generated by the Sine-Sine-map. In [20], an image cryptosystem based on dynamic DNA encryption and chaos has been suggested. In [21], the authors have introduced a 2D chaotic map generated by connecting the Henon map and Sine map. In [22], a method for image encryption using fractional Fourier transform, DNA sequence operation, and chaos theory have been developed. In [23], the authors have put forward a cryptosystem based on a 2D chaotic map derived from a Sine map, Chebyshev map, and a linear function. In [24], a 4D dimensional memristive chaotic system has been constructed based on Liu's chaotic system by introducing a flux-controlled memristor model. In [25], a multiple-image encryption scheme based on DNA sequence and image matrix has been studied to provide fast and secure indexes. In [26], the authors have been propounded an image encryption algorithm based on DNA sequence operations to improve data security. In [27], the authors have implemented a Pareto-optimal image encryption scheme using a coupled map lattice chaos function and DNA combination. When those suggested schemes are surveyed in the view of encryption performance, it is seen that they have their own strong and weak sides across the encrypting operations give successful results for particular analysis approaches which were performed. The keyspace results in [19,23,24,27] comparatively seem the lowest within the analysis carried out in [8–13,15–17,20,21]. The information entropies can be assessed as better [12,13,21,24,25], moderate [20,22,23] and worse [15,18,27] in comparison among each other. While the correlation results in [13–16,19–21,25] are lower, those in [12,17,18,23,24,26] are moderate and ones in [22,27] are higher. The resistances of the scheme counter the differential attacks can be sorted from the strongest to the weakest as [13,20,23], [14,17,21,24], and [12,15,16,18,19,22,25–27]. The scheme in [19] appears better against cropping and noise attacks among the available results [14–21,23,24]. However, for the above methods, by using several attack methods such as differential attack analysis, cropping attack analysis, and noise attack analysis, encrypted images still can be exploited. Hence, a more secure image encryption method to improve security is necessary.

In this study, an image encryption scheme based on a novel chaotic map and key generation is proposed. The scheme is built on a logarithmic chaotic map and public key generated through a deep convolution neural network (CNN). The new contributions of the presented study can be emphasized as follows:

- A key generation method through a designed CNN model based on VGG16 architecture is proposed.
- A chaotic map referred to as log-map is introduced for hyperchaotic sequence.
- An operation so-called bit reversion is used for the manipulation process.

First, a main public key is obtained with XOR operation between a secret key and the public key generated through the deep CNN model. Then, four initial values and four control parameters are produced to be used in the chaotic log-maps, a part of the encrypting operations such as permutation, DNA encoding, diffusion, and bit reversion, respectively. Afterward, the performance of the encryption scheme for the well-known images is comparatively analyzed across the reliable metrics regarding visual and numerical views. Based on observed results, the proposed scheme, thanks to having a sensitive key, hyperchaotic log-map, and impactful bit reversion operation, outperforms state-of-the-art techniques.

## 2. The proposed image encryption scheme based on log-map

In the proposed image encryption scheme whose block diagram is given in Fig. 1, a new chaotic map, a feature extraction-based public code, and bit reversion is introduced through the scheme. The ciphertext image is obtained through four operation stages: permutation, DNA coding, diffusion, and bit reversion. The main key is generated by XOR operation between the public key and a secret key. The chaotic sequence is obtained using the log-map by serving the main key as the control parameter and initial value to the map. The four operation stages are then processed according to the chaotic sequence, and the image is encoded with the DNA rules.

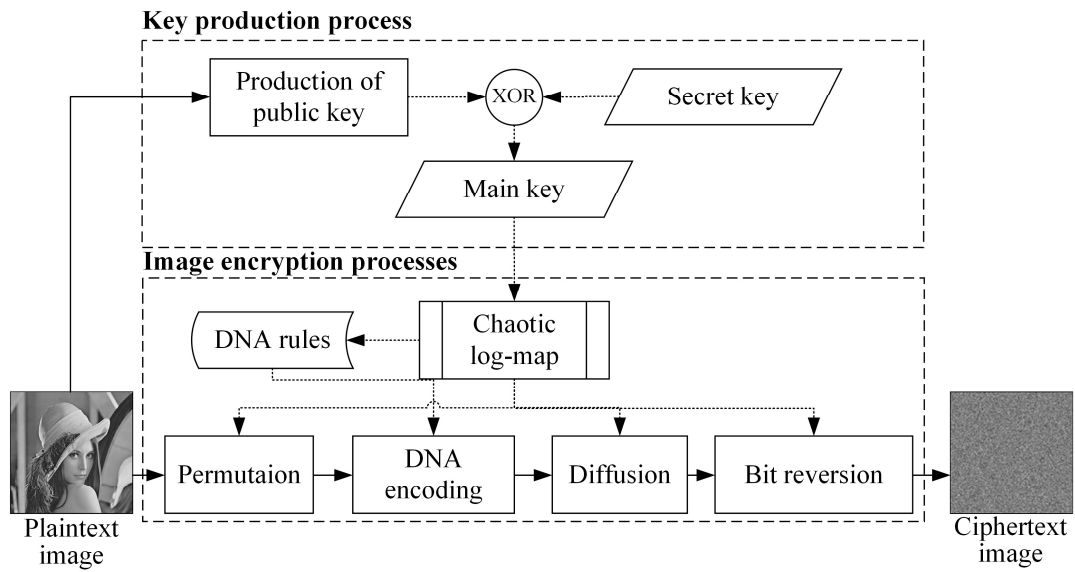

**Fig. 1.** An overview of the processes of the proposed image encryption scheme.

### 2.1 Generation of public code based on feature extraction using deep CNN and the main key

This section generates a public key via a deep CNN model [28] depicted in Fig. 1 by employing VGG16 architecture with the pre-trained network having an Image-Net dataset with 1000 categories. The model is implemented using Keras library's Tensorflow backend [29]. The CNN model consists of five layers with convolution and pooling processes. The dimension of the extracted features for each image is $7 \times 7$ with 512 channels. Therefore, the number of features is $7 \times 7 \times 512 = 25088$ that converted to a one-dimensional vector using a flatten layer added to the model. After the convolution, there are two additional layers and pooling layers referred to as Dense-1 and Dense-2 functioning for dimension reduction. The dense layers' parameters are randomly initiated with Glorot-Uniform distribution that adjusts the network's initial weights [30]. By random initialization, the model generates a different public key for the same image at each time. The variance of the dense layers' outputs is close to the input variance. Sigmoid is used as the activation function for every dense layer. Image features herewith get a value between 0 and 1 before being converted to binary. Binarization is achieved by comparing the output of Dense-2 with 0.5.

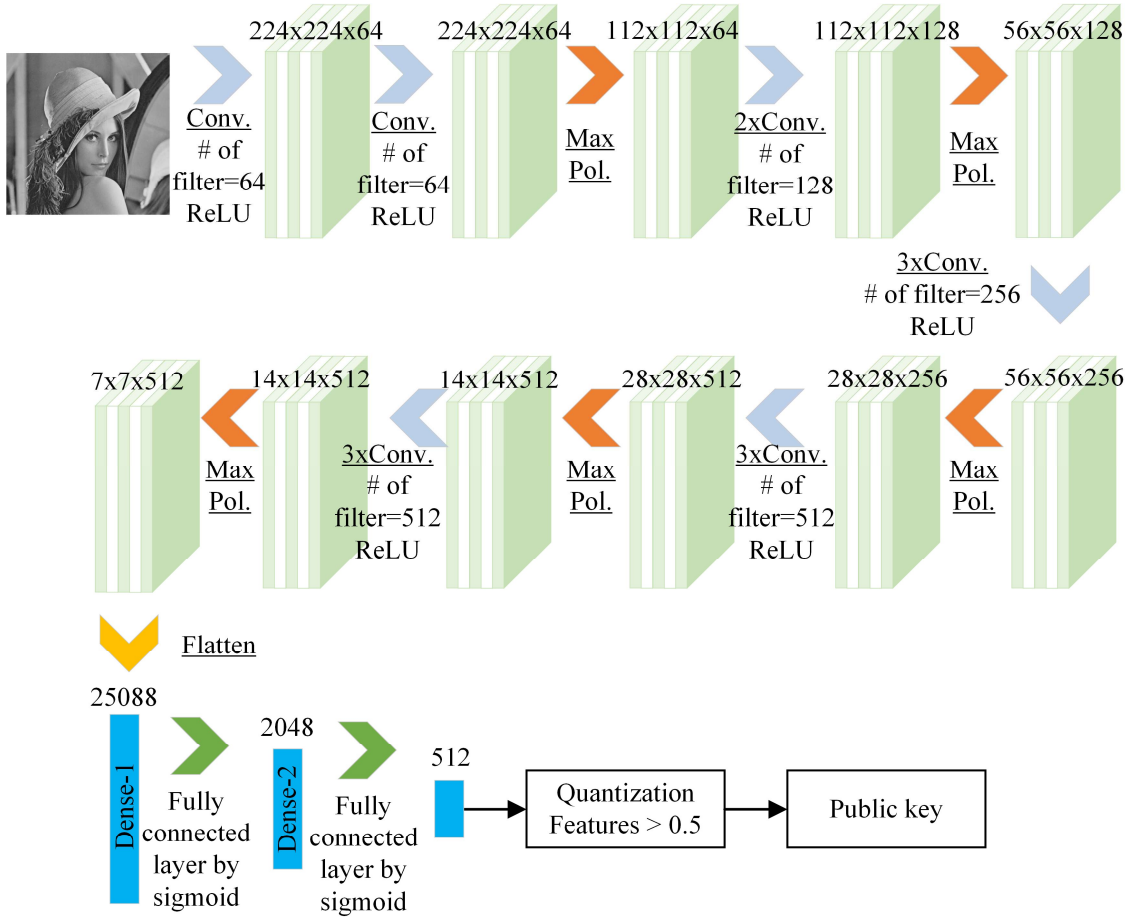

**Fig. 2.** The proposed deep CNN model and key generation procedure.

## 2.2 The proposed chaotic map: log-map

Chaotic maps based on various logistic equations are frequently employed in the image encryption schemes. The chaotic maps are used to generate a diverse sequence in accordance with the control parameter and initial value. The pixels of the image to be encrypted are herewith scrambled and manipulated through the generated sequence. To explain the working principle of the chaotic maps, a conventional 1D chaotic map (Eq. 1) and its Lyapunov exponent (LE) (Eq. 2) can be given as follows:

$$v_{i+1} = uv_i(1 - v_i), \ v_i \in (0,1) \tag{1}$$

$$\text{LE} = \lim_{n \to \infty} \frac{1}{n} \sum_{i=1}^{n-1} \ln|u - 2uv_i| \tag{2}$$

Here, $u \in [0, 4]$ is the control parameter (growing rate), $v_i$ is the initial value, and $n$ is the number of iterations for a specific value of the control parameter.

Bifurcation diagram, a plot of output values versus the map's control parameter, shows a multi-diverse solution suddenly appears while the control parameter changes. These solutions are also called bifurcation points [31]. Moreover, the LE, the first mentioned in [32], is exploited to evaluate a chaotic map's performance regarding the system's predictability and sensitivity to the control parameter and initial value. Besides, the LE should be positive; the higher LE is, the better the chaotic characteristic shows.

The bifurcation diagram of the conventional chaotic map provided Eq. (1) and its LE plot given in Eq. (2) are illustrated in Fig. 3(a) and (b), respectively. It is seen that the LE is positive only for $u \in [3.57, 4)$ and hence the conventional chaotic map does not behave chaotically except for this range. Because the chaotic range of a map should be larger for a more secure image encryption scheme, we propose a novel chaotic map particularly denoted log-map in Eq. (3) with a convenient LE equation in Eq. (4).

$$v_{i+1} = \mathrm{mod}\big((u + e)\ln v_i, 1\big), \;\; v_i \in (0,1) \tag{3}$$

$$\text{LE} = \lim_{n \to \infty} \frac{1}{n} \sum_{i=1}^{n-1} \ln\left(\frac{u+e}{v_i}\right), \;\; (u+e)\ln v_i > \mathrm{floor}\big((u+e)\ln v_i\big) \tag{4}$$

Here, $u \in [0, \infty)$ is the control parameter, $v_i$ is the initial value, $n$ is the number of iterations for a specific value of the control parameter, and the floor is the greatest integer function.

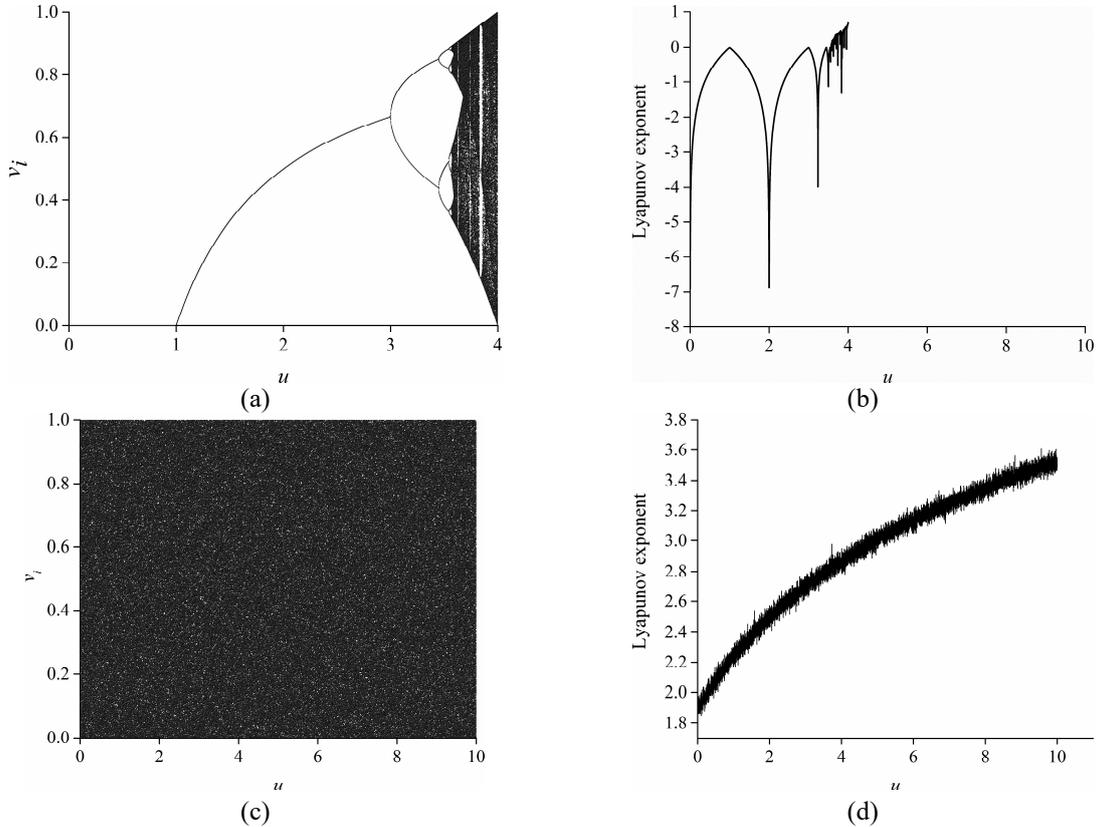

**Fig. 3.** Bifurcation sequence and LE plot (a) Bifurcation diagram of the chaotic map, (b) LE of the chaotic map, (c) Bifurcation diagram of the log-map, (d) LE of the log-map

The bifurcation diagram of the log-map and the relevant LE are plotted in Fig. 3(c) and (d), respectively. The figures manifest that the LE value is positive for all nonnegative $u \in \mathbb{R}$. The log-map produces a hyperchaotic map with higher ergodicity and diversity for a larger range of the control parameter besides with a higher LE value.

### 2.3 DNA rules and encoding

DNA, which is indeed a biological concept, is a molecule that carrying out genetic codes for surviving and continuing their generations. DNA consists of four nucleotides, adenine (A), thymine (T), cytosine (C), and guanine (G). In DNA rules, a binary number with two digits stands for each nucleotide such as A-00, C-01, G-10, and T-11. Therefore, a code with a variety of 24 combinations can be performed by using this rule. As shown in Table 1, these combinations can be reduced to eight type encoding rules according to Watson-Crick complementary rule [33]. Every grayscale pixel is illustrated with 8 bits. It corresponds to a DNA array with four nucleotides. For instance, the 167-pixel value is defined as a 10100111 binary array and it can be encoded as 00001110 by utilizing rule-4, which is also AATG in terms of DNA nucleotide. Thanks to this simple and effective encoding strategy, the DNA rule is successfully exploited to achieve a secure encryption scheme by manipulating the pixel values of the image.

**Table 1.** DNA encoding rules

| Rule   | 1 | 2 | 3 | 4 | 5 | 6 | 7 | 8 |
|--------|---|---|---|---|---|---|---|---|
| 00 (A) | A | A | C | G | C | G | T | T |
| 01 (C) | C | G | A | A | T | T | C | G |
| 10 (G) | G | C | T | T | A | A | G | C |
| 11 (T) | T | T | G | C | G | C | A | A |

## 3. Processes of the proposed encryption scheme

This section presents a novel image encryption scheme based on the log-map and provides some of the basic notions. The proposed encryption scheme is seriatim processed through four encrypting operations called a permutation, DNA encoding, diffusion, and bit reversion.

Throughout this paper, let $A \coloneqq [a_{ij}]_{m \times n}$ be a gray-scale image matrix such that $i \in \{1,2,\ldots,m\}$, $j \in \{1,2,\ldots,n\}$, $a_{ij}$ is an unsigned integer number, $0 \le a_{ij} \le 255$, and $\text{floor}(A) \coloneqq [\text{floor}(a_{ij})]$. Moreover, let $\hat{a}_{ij}$ denote a binary form of $a_{ij}$. Then, $\hat{A} \coloneqq [\hat{a}_{ij}]_{m \times n}$ referred to as the converted matrix of $A$ to base 2.

**Definition 2.1** Let $A \coloneqq [a_{ij}]_{m \times n}$ and $B \coloneqq [b_{ij}]_{m \times n}$ be two image matrices. Then, $C$ is called the binary sum of $A$ and $B$ and is denoted by $C = A \oplus B$, if $\hat{C} \coloneqq mod(\hat{A} + \hat{B}, 2)$.

### 3.1 Obtaining the initial values and control parameters

In the encryption scheme, firstly, the public key is generated using the deep CNN following the procedure expressed in Section 2.1. Then, the initial value $v$ and the control parameter $u$ are obtained to produce chaotic sequences through the log-map that employs the public and secret keys. The chaotic sequences are utilized in the encrypting operations: permutation, DNA encoding, diffusion, and bit reversion. The production of the initial values and control parameters is shown in Algorithm 1, and even elaborately depicted in Fig. 4 through an illustrative example of the Lena image. The algorithm manages the generation of the public key referred to as binary matrix $C$ via deep CNN in Steps 1-3. Once a secret key matrix $D$ is constructed in Step 4, the main key matrix $E$ is acquired by XOR operation between the public and secret keys in Step 5, and then the main key is divided into eight matrix groups, each of which used to produce the initial values and control parameters. In Step 6, the columns of each matrix group in self have imposed in turn a series of transaction $\text{mod}(\text{sum}(E, 2), 2)$ then the results of each group $8 \times 1$ are combined to be $F$ matrix size $8 \times 8$. In Step 7, the binary matrix $F$ is afterward converted to a decimal matrix $S$. Finally, the initial values $V := [v_1 \ v_2 \ v_3 \ v_4]$ and control parameters $U := [u_1 \ u_2 \ u_3 \ u_4]$ are respectively obtained by transactions $\frac{s_{1i}}{256}$ and $\frac{s_{1,i+4}}{256} + \text{mod}(s_{1,i+4}, 10)$ in Steps 8 and 9.

**Algorithm 1:** Obtaining of initial values and control parameters

**Step 1.** Read an image matrix $A$

**Step 2.** Obtain a fuzzy row matrix $B := [b_{1p}]_{1 \times 512}$ by applying deep CNN to $A$

**Step 3.** Obtain the public key $C := [c_{1p}]_{1 \times 512}$ defined by $c_{1p} := \begin{cases} 1, & b_{1l} > 0.5 \\ 0, & \text{otherwise} \end{cases}$

**Step 4.** Construct a secret key $D := [d_{1p}]_{1 \times 512}$ being a binary row matrix

**Step 5.** Evaluate the main key $E := C$ XOR $D$ and reshape $E$ to $8 \times 8 \times 8$ in shape

**Step 6.** Compute $F := \text{mod}(\text{sum}(E, 2), 2)$ and then reshape $F$ to $8 \times 8$ in shape where $\text{sum}(E, 2)$ means the sum of the rows of each submatrix in shape $8 \times 8$ of $E$

**Step 7.** Figure out $S := [128 \ 64 \ 32 \ 16 \ 8 \ 4 \ 2 \ 1]F$

**Step 8.** Obtain the initial values $v_i \leftarrow \frac{s_{1i}}{256}, i \in \{1,2,3,4\}$

**Step 9.** Obtain the control parameters $u_i \leftarrow \frac{s_{1,i+4}}{256} + \text{mod}(s_{1,i+4}, 10), i \in \{1,2,3,4\}$

| | |
|---|---|
| Public Key $C$ | 0 1 1 1 1 1 1 1 … 0 0 0 1 |
| Secret Key $D$ | 0 1 0 1 0 1 1 0 … 0 0 0 1 |
| Main Key $E := \text{mod}(C + D, 2)$ | 0 0 1 0 1 0 0 1 … 0 0 0 0 |
| Reshaped $E$ has order $8 \times 8 \times 8$ | (matrices shown) |
| $F := \text{mod}(\text{sum}(E, 2), 2)$ has order $8 \times 1 \times 8$ | (column vectors shown) |
| Reshaped $F$ has order $8 \times 8$ | (matrix shown) |
| $S := [128 \ 64 \ 32 \ 16 \ 8 \ 4 \ 2 \ 1]F$ | [210 37 163 66 189 110 87 138] |
| Initial Values $V := [v_1 \ v_2 \ v_3 \ v_4]$ | [210 37 163 66]/256 |
| Control Parameter $U := [u_1 \ u_2 \ u_3 \ u_4]$s | [189 110 87 138]/256 + mod([189 110 87 138],10) |

**Fig. 4.** The procedure to obtain the initial values and control parameters for processes of the encryption scheme

## 3.2 The processes of the encryption scheme

The proposed scheme conducts the encryption of an image through Algorithm 2 consisting of four encrypting operations: permutation, DNA encoding, diffusion, and bit reversion. The processes of the encryption scheme are presented in Fig. 5 with an illustrative example for a 5x5 pixel sample of the Lena image. The pixels are scrambled and manipulated using the operations according to the chaotic sequence produced by the log-map. The obtained initial value and control parameter pairs are herein utilized for each operation to obtain chaotic sequences in Steps 1, 4, 8, and 11, i.e. $v_1$ and $u_1$ pair is for the permutation, $v_2$ and $u_2$ pair is for the DNA encoding, $v_3$ and $u_3$ pair is for the diffusion, $v_4$ and $u_4$ pair is for the bit reversion. In the permutation (Steps 2 and 3), the position of every pixel is transferred with regard to the ascending order of the chaotic sequence. In this way, the position of every pixel is scrambled under the management of the main key and log-map. The pixel values are seriatim manipulated in the remaining three operations. In the DNA encoding (Steps 5-7), the chaotic sequence is exploited to determine the DNA rule from Table 1. The tonal value of every pixel is changed in accordance with the determined DNA rule. In diffusion (Steps 9 and 10), the encoded pixels are undergone an XOR operation with a matrix $Y_3$ attained via the chaotic sequence. The last process, which is the bit reversion, is a new operation proposed in this study to improve the security further. In the bit reversion (Steps 12-14), diffused pixels are incurred an XOR operation with a matrix $Z$ acquired using the reverted bits of a matrix $Y_4$ obtained by the chaotic sequence. The outcome matrix $A_4$ of the bit reversion operation is thus the final ciphertext image.

---

**Algorithm 2:** The steps for the operations of the image encryption scheme

**Permutation**

Step 1. Compute the first chaotic sequence $X_1 := [x^1_{1r}]_{1 \times mn}$ by using the initial value $v_1$, the control parameter $u_1$, and the log-map (see Eq. (3))

Step 2. Obtain $Y_1 := [y^1_{1r}]_{1 \times mn}$ by sorting $X_1$ in ascending order

Step 3. Compute a Permutation matrix $A_1 := [a^1_{kl}]_{m \times n}$ defined by $a^1_{kl} := a_{ij}$ such that $x^1_{1,n(i-1)+j} = y^1_{1,n(k-1)+l}$

**DNA Encoding**

Step 4. Compute the second chaotic sequence $X_2 := [x^2_{1r}]_{1 \times mn}$ by using the initial value $v_2$, the control parameter $u_2$, and the log-map

Step 5. Evaluate the row matrix $Y_2 := [y^2_{1r}]_{1 \times mn}$ defined by $Y_2 := 1 + \text{floor}(8X_2)$

Step 6. For all $k$ and $l$, convert $a^1_{kl}$ to base-2 and rearrange these converted entries according to the rules that correspond to the values $y_{1r}$ (see Table 1)

Step 7. For all $k$ and $l$, convert the rearranged entries to base-10 and construct the DNA Encoding matrix $A_2 := [a^2_{kl}]_{m \times n}$

**Diffusion**

Step 8. Compute the third chaotic sequence $X_3 := [x^3_{1r}]_{1 \times mn}$ by using the initial value $v_3$, the control parameter $u_3$, and the log-map

Step 9. Evaluate the row matrix $Y_3 := [y^3_{1r}]_{1 \times mn}$ defined by $Y_3 := \text{floor}(256X_3)$ and then reshape $Y_3$ to $m \times n$ in shape

Step 10. Compute the Diffusion matrix $A_3 := [a^3_{kl}]_{m \times n}$ defined by $A_3 := A_2 \oplus Y_3$

**Bit Reversion**

Step 11. Compute the third chaotic sequence $X_4 := [x^4_{1r}]_{1 \times mn}$ by using the initial value $v_4$, the control parameter $u_4$, and the log-map

Step 12. Evaluate the row matrix $Y_4 := [y^4_{1r}]_{1 \times mn}$ defined by $Y_4 := \text{floor}(256X_4)$

Step 13. For all $r$, retype $\hat{y}^4_{1r}$ in reverse order and set to $\hat{z}_{1r}$. Then, reshape $Z := [z_{1r}]_{1 \times mn}$ to $m \times n$ in shape

Step 14. Compute the Bit Reversion matrix $A_4 := [a^4_{kl}]_{m \times n}$ defined by $A_4 := A_3 \oplus Z$

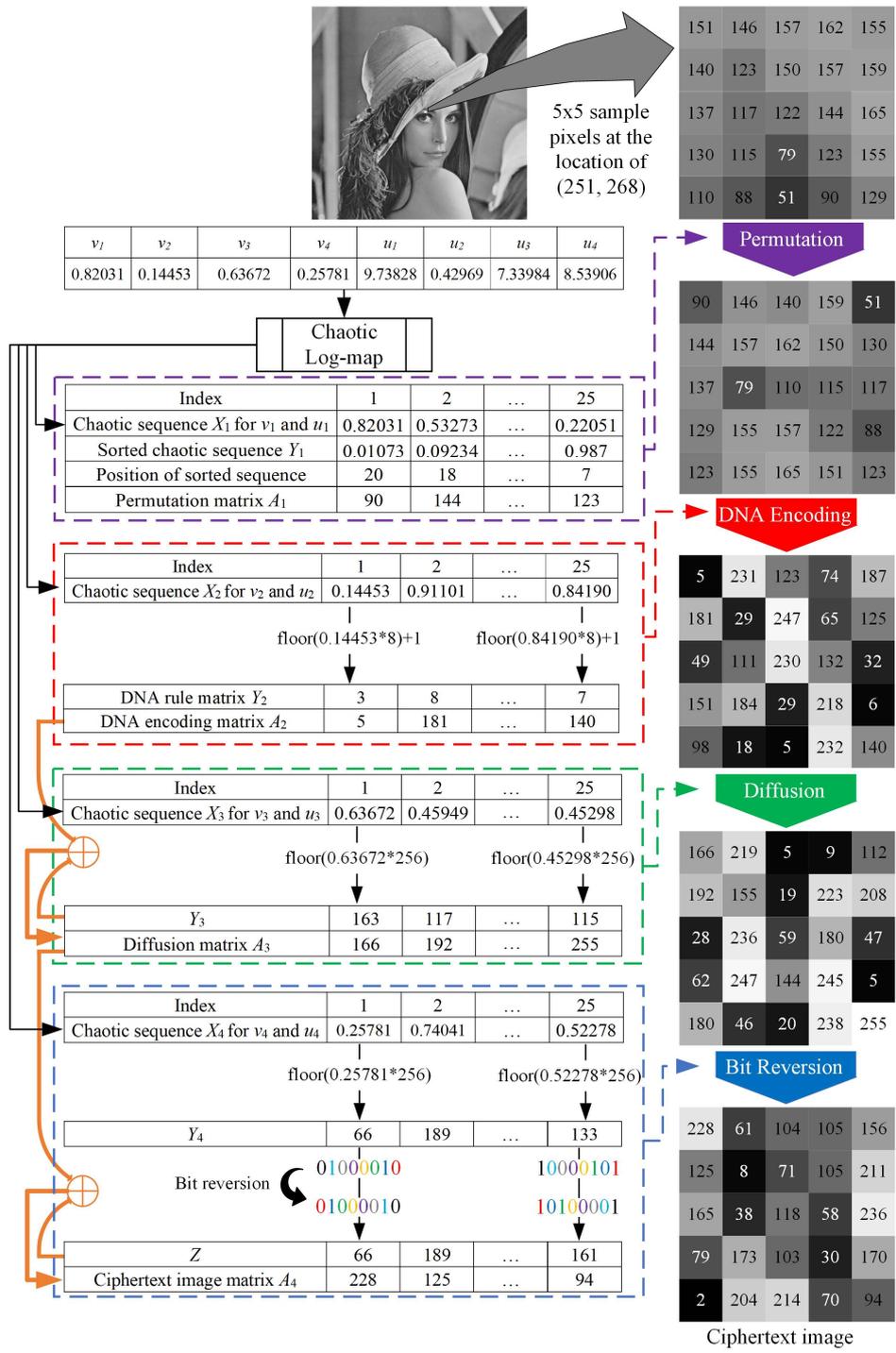

**Fig. 5.** The processes of the proposed encryption scheme with an illustrative example

## 4. Comparative experimental results and performance analyses

The main aim of an encryption scheme is to improve its proof against potential cyber threats through network attaches such as denial-of-service, man-in-the-middle, and phishing. There are various cryptanalysis methods such as keyspace, key sensitivity, entropy, histogram, correlation, differential attack, noisy attack, and cropping attack to evaluate the security of an encryption scheme by simulating some cyber threats [11]. This cryptanalysis methods are performed on the well-known images with size $512 \times 512$. The classification results are carried out via MATLAB R2018b and Python 3.5 running on a workstation with I(R) Xeon(R) CPU

E5-1620 v4 @ 3.5 GHz, and 64 GB RAM. The scores of entropy, correlation, and differential attack are obtained on average for 200 run times. The outcomes of the proposed encryption scheme are even compared with some the-state-of-the-art results reported elsewhere [12,13,37,38,15,18,21,23,24,34–36].

### 4.1 Keyspace analysis

Brute-force is a type of cyberattack based on predicting the key by trying numerous possible passwords or passphrases. An image encrypted with a short key is inherently vulnerable to this attack in time. In case the key is longer, it would resist for a long time. Therefore, it would be impossible to guess the key if it has the proper length. Keyspace analysis is utilized for testing the proof capability of the Brute-force attacks. According to this analysis, a key with longer than $2^{100}$ is considered for high security encryption [39]. In our scheme, we even propose an approach depending on deep CNN for generation SHA 512. Based on this key, eight floating numbers with $10^{15}$ precision is obtained to be used as the initial values and control parameters of the encrypting operations. Therefore, the keyspace of the proposed scheme is $10^{15 \times 8} = 10^{120} \cong 2^{398}$ which is much higher than that of $2^{100}$.

### 4.2 Key sensitivity analysis

A secure encryption scheme should also be highly sensitive to the key. In other words, a slight change in the key must result in a significant variation in the image. Several tests can be performed in order to appreciate the key sensitivity. To this end, five secret keys are given in Table 2 contains an original key and its versions obtained by slightly changing the first digit. The ciphertext images encrypted via those keys are illustrated in Fig. 6, and their differential images are presented to visually observe the number of pixels with the same tonal values. The same pixels with the same tonal values would seem black due to their zero difference values. As can be seen from Fig. 6(d), (f), (h), and (j), there does not seem any black region in those images.

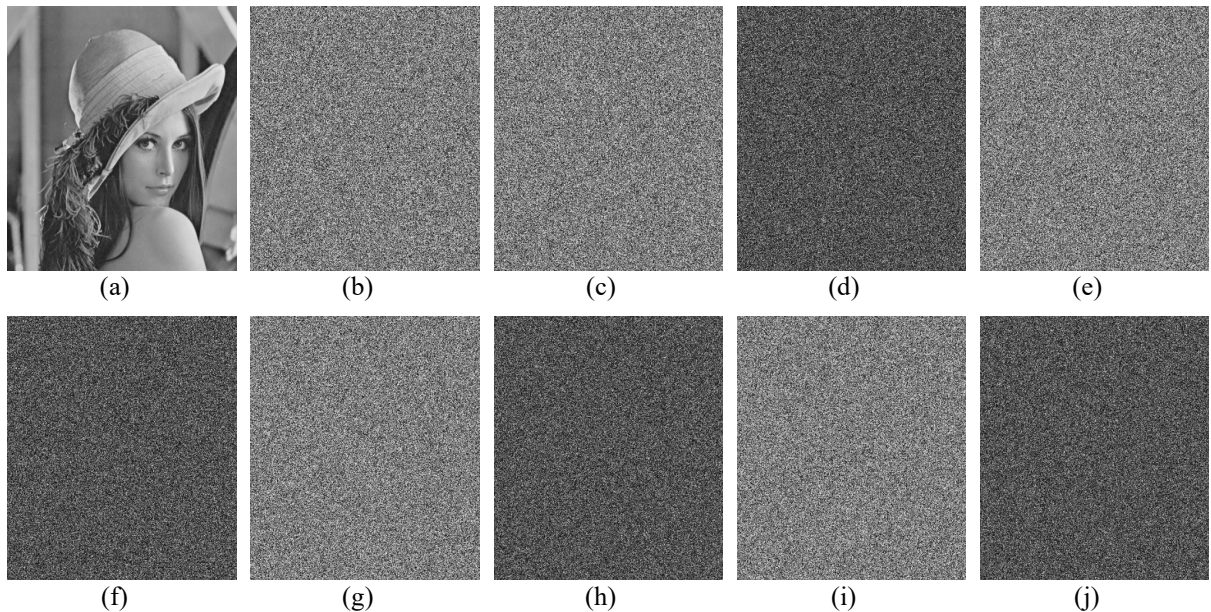

**Fig. 6.** Key sensitivity test: (a) plain image; (b) ciphertext with key 1; (c) ciphertext with key 2; (d) differential image between (b) and (c); (e) ciphertext with key 3; (f) differential image between (b) and (e); (g) ciphertext with key 4; (h) differential image between (b) and (g); (i) ciphertext with key 5; (j) differential image between (b) and (i).

The distinctive level of the differential images with that of the original key in Fig. 6 is calculated to evaluate the key sensitivity numerically in Table 3. The distinctive levels are determined 99.6250%, 99.6074%, 99.6265% and 99.6067% respectively for Key 2, 3, 4 and 5. Moreover, the mean of these levels is 99.6164%. It is evident from the results that the proposed encryption scheme is very sensitive to the key thanks to its diversity performance.

**Table 2.** A secret key and slightly changed versions of this key by altering the first digit.

| Key | SHA-512 |
| --- | --- |
| Key 1 (original) | 010101100011100101111001001001000100001000100110010001010010100…1 |
| Key 2 (changed) | 01**1**00110001110010111100100100100010000100010011001000101001010010100…1 |
| Key 3 (changed) | 011**1**0110001110010111100100100100010000100010011001000101001010010100…1 |
| Key 4 (changed) | **1**00001100011100101111001001001000100001000100110010001010010100101 00…1 |
| Key 5 (changed) | 100**1**0110001110010111100100100100010000100010011001000101001010010100…1 |

**Table 3.** Differences between cipher images produced by slightly different keys.

| Figures | Encryption keys | Difference with 1(b) (%) |
| --- | --- | --- |
| 1(c) | Key 2 | 99.6250 |
| 1(e) | Key 3 | 99.6074 |
| 1(g) | Key 4 | 99.6265 |
| 1(i) | Key 5 | 99.6067 |
| Mean | | 99.6164 |

Another test for measuring the key sensitivity is to decrypt the ciphertext image that is encrypted via the slightly changed keys. In this wise, it is aimed to visually examine the decrypted ciphertext images with the changed keys so that whether they involve any information related to the plaintext image. Fig. 7 comparatively shows the decrypted images together with that of the ciphertext image with the original key in Fig. 7(b). While the cyphertext image with the original key is exactly decrypted as the same plaintext image, the other images do not contain any texture from the plaintext image.

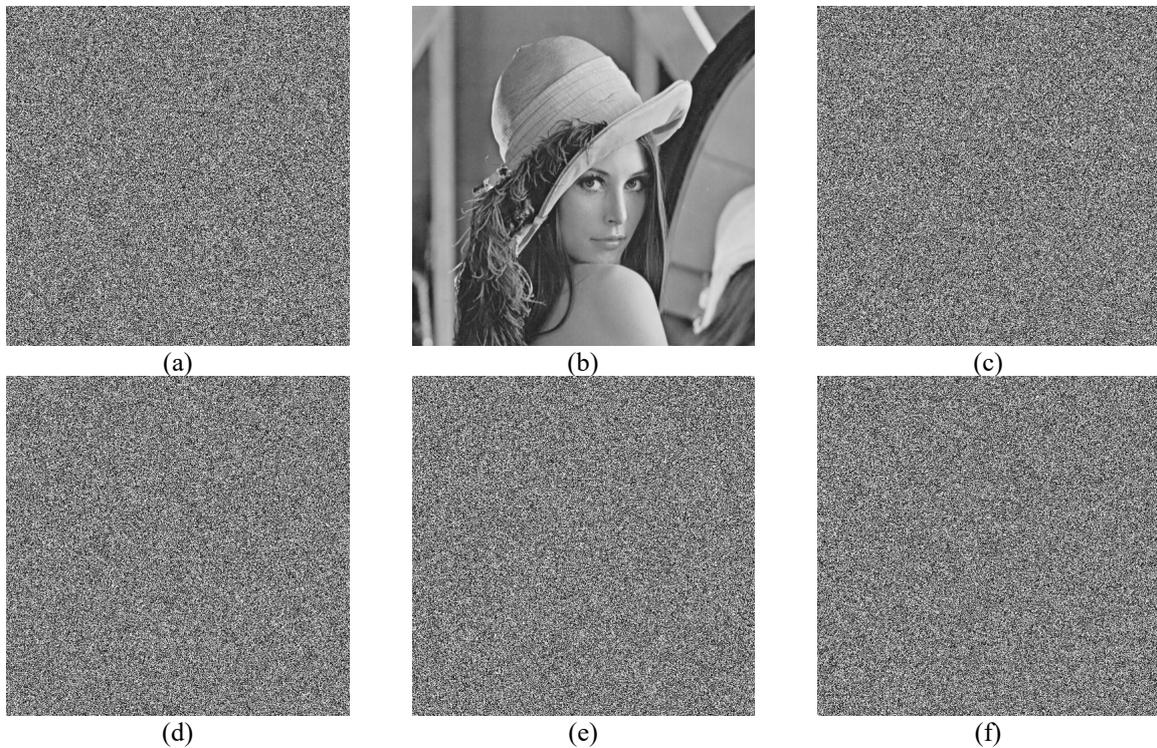

(a)  (b)  (c)
(d)  (e)  (f)

**Fig. 7.** Key sensitivity in the second case: (a) cipher image with *key 1*; (b) decipher image with *key 1*; (c) decipher image with *key 2*; (d) cipher image with *key 3*; (e) decipher image with *key 4*; (f) decipher image with *key 5*.

## 4.3 Histogram Analysis

A histogram presents a graphical distribution of the pixel value's number versus every tonal value. Therefore, it allows investigating the uniformity of the image pixels. The manipulation performance of the encryption scheme can thus be evaluated through the histogram analysis. The more uniform histogram, the better the manipulation performance. The histograms of the well-known plaintext and their ciphertext images Lena, Cameraman, Baboon, Peppers, Airplane, and Barbara are illustrated in Fig. 8. It can be clearly seen that the proposed image encryption scheme uniformly manipulates the tonal value of the images, implying that it is not able to extract any information from the ciphertext images regarding the plaintext images. Therefore, the image encryption scheme can resist statistical attacks.

In order to analyze further the distribution of the pixels' tonal values, variance and $\chi^2$ tests of the histogram are calculated. For a grayscale image, variance and $\chi^2$ tests can be computed as follows:

$$var(X) = \frac{1}{n^2} \sum_{i=0}^{n} \sum_{j=1}^{n} \frac{1}{2}(x_i - x_j)^2 \quad (5)$$

$$\chi^2 = \sum_{i=0}^{255} \frac{(n_i - n/256)^2}{n/256} \quad (6)$$

where $n_i$ is the repetition frequency of a tonal value $i$ and $n$ is the number of total pixels. $n/256$ is hence the expected repetition frequency for each tonal value. $X=\{x_1, x_2, ..., x_{256}\}$ is the vector of the histogram's tonal values. $x_i$ and $x_j$ are the numbers of pixels whose gray values are equal to $i$ and $j$, respectively. For desiring high uniformity, the variance should be lowered as much as possible. On the other side, $\chi^2(0.05; 255)$ should be lower than 293.25 for passing $\chi^2$ test with 0.05, which is the significant level [40]. The results of variance and $\chi^2$ test are tabulated in Table 4 for the images under the histogram analysis in Fig. 8. The proposed image encryption scheme is therefore verified in terms of the results in the table for all the images under the analysis.

**Table 4.** Variance and $\chi^2$ test results of the images under the histogram analysis

| Test | Image | Lena | Cameraman | Peppers | Baboon | Barbara | Airplane |
|---|---|---|---|---|---|---|---|
| var | Plaintext | 6333788.75 | 1674120.58 | 2196605.10 | 845463.33 | 3821955.00 | 2832714.39 |
| | Ciphertext | 993.10 | 997.77 | 1007.18 | 988.73 | 980.32 | 1006.91 |
| $\chi^2$ | Plaintext | 158344.71 | 418530.14 | 549151.27 | 211365.83 | 95548.87 | 708178.59 |
| | Ciphertest | 247.30 | 248.46 | 250.81 | 252.81 | 245.60 | 250.74 |

## 4.4 Information entropy analysis

Information entropy is the most applied analysis to measure the uncertainty and disorderliness of a ciphertext image [4]. Therefore, it reflects the manipulation performance of an image encryption scheme. Information entropy of an image can be computed by the following equation.

$$H(x) = \sum_{i=0}^{2^n-1} p(x_i) \log_2 \frac{1}{p(x_i)} \quad (7)$$

Here, $x$ is the information source. The probability of $x_i$ can be represented by $p(x_i)$ and $2^n$ referring to the overall states.

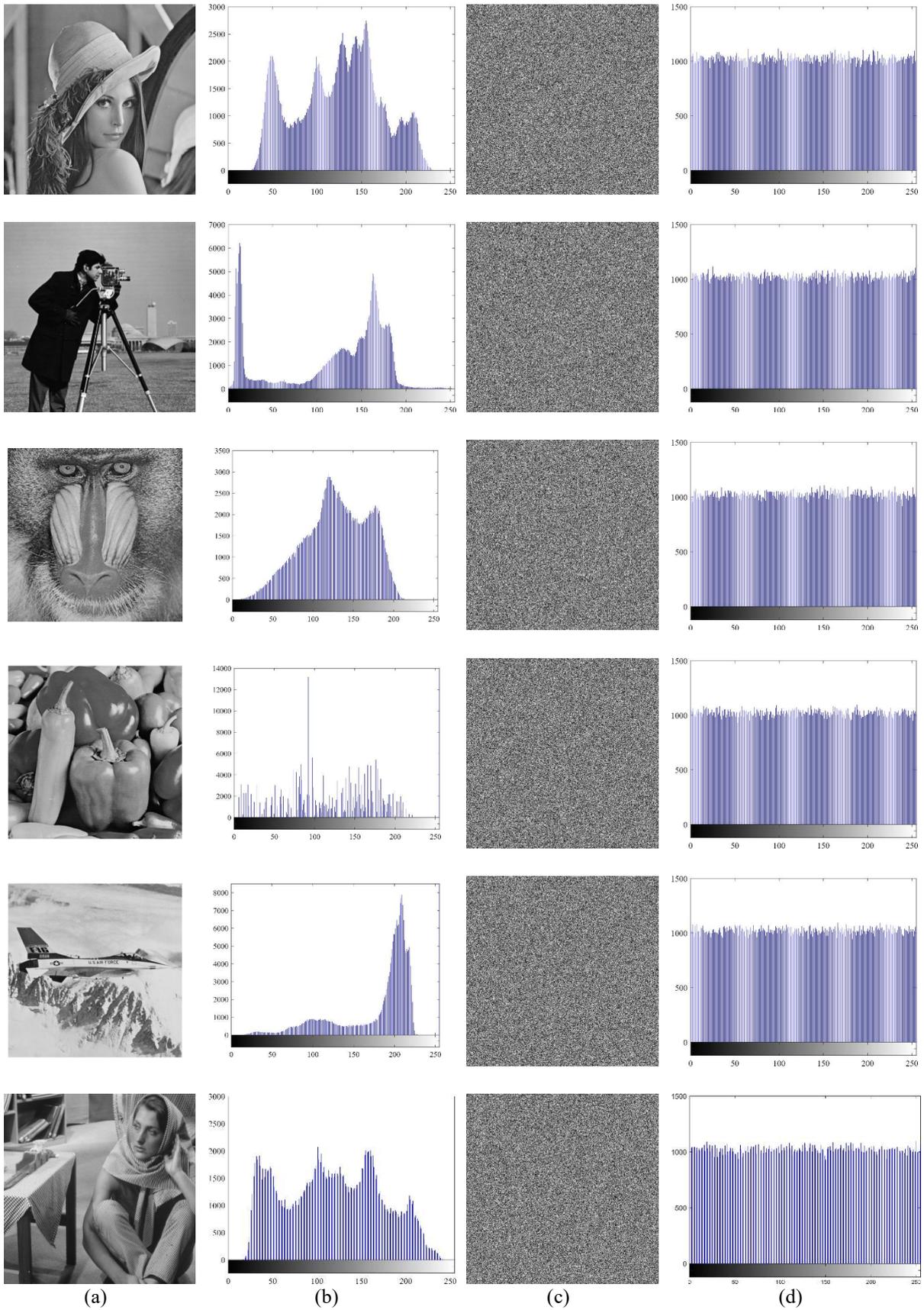

**Fig. 8.** Histograms of the well-known images: (a) the plaintext images, (b) histograms of the plaintext images, (c) the ciphertext images, (d) the histograms of ciphertext images.

The maximum entropy of an encrypted image can theoretically be 8 for a grayscale image. Therefore, the uncertainty and disorderliness performance of an image encryption scheme are evaluated being how close to this theoretical value. The information entropy of the under-analysis images encrypted through the proposed scheme is given in Table 5. They are compared with the available state-of-the-art results [12,13,15,18,21,23,24] for Lena, Cameraman, Peppers, and Barbara ciphertext images in Table 6. As seen from Table 5, all information entropy values of the ciphertext images are closely near to 8. Moreover, the proposed scheme encrypts images with the best information entropy of 7.9994 among the other suggested schemes [12,13,15,18,21,23,24]. These cryptoanalysis results mean that the proposed scheme assures the most assured images against cyberattacks.

**Table 5.** Information entropy of the plaintext and ciphertext images encrypted through the proposed scheme.

| Image | Lena | Cameraman | Baboon | Peppers | Barbara | Airplane |
|---|---|---|---|---|---|---|
| Plaintext image | 7.4455 | 7.0479 | 7.2925 | 6.7624 | 7.6321 | 6.7135 |
| Ciphertext image | 7.9994 | 7.9994 | 7.9994 | 7.9994 | 7.9993 | 7.9994 |

**Table 6.** The comparative information entropy of several ciphertext images of several methods.

| Ciphertext image | ReF. [12] | ReF. [13] | ReF. [15] | ReF. [18] | ReF.[21] | ReF. [23] | ReF. [24] | Proposed scheme |
|---|---|---|---|---|---|---|---|---|
| Lena | 7.9993 | 7.9993 | 7.9975 | 7.9982 | 7.9994 | 7.9970 | 7.9994 | **7.9994** |
| Cameraman | - | - | - | - | 7.9993 | 7.9973 | 7.9970 | **7.9994** |
| Peppers | - | 7.9994 | - | - | 7.9993 | 7.9969 | - | **7.9994** |
| Barbara | 7.9992 | - | 7.9985 | 7.9981 | - | - | - | **7.9993** |

## 4.5 Correlation analysis

High correlation inherently exists among the adjacent pixels of a plaintext image. However, a secure image encryption scheme must alleviate the correlation by introducing an effective permutation operation. The correlation coefficient of an image can be computed in horizontal, vertical, and diagonal directions using the following equation.

$$r_{xy} = \frac{E[x - E(x)][y - E(y)]}{\sqrt{D(x)}\sqrt{D(y)}} \qquad (8)$$

where the two auxiliary equations are $E(x) = \frac{1}{N}\sum_{i=1}^{N} x_i$ and $D(x) = \frac{1}{N}\sum_{i=1}^{N}[(x_i - E(x)]^2$. Here, $x_i$ and $y_i$ are tonal values of $i$-th pair of the selected adjacent pixels, and $N$ represents the number of the pixel samples. For the correlation analysis performed in this study, $N = 3000$ pixel samples are randomly selected from the ciphertext image.

The correlation coefficients of the under-analysis images encrypted through the proposed scheme are given in Table 6, and they are also compared with the available correlation coefficients in the literature for Lena, Cameraman, and Peppers ciphertext images in Table 8 [12,13,15,18,21,23,24]. From Table 7, the proposed image encryption scheme reduces the correlation coefficients to be very close to zero. Furthermore, it outperforms the other schemes suggested in the literature in terms of the correlation coefficients tabulated in Table 8.

**Table 7.** Correlation coefficients of the ciphertext images under-analysis.

| Direction | Lena | Cameraman | Baboon | Peppers | Barbara | Barbara | Airplane |
|---|---|---|---|---|---|---|---|
| horizontal | $-29 \times 10^{-5}$ | $27 \times 10^{-5}$ | $52 \times 10^{-7}$ | $-80 \times 10^{-6}$ | $86 \times 10^{-6}$ | $86 \times 10^{-6}$ | $47 \times 10^{-6}$ |
| vertical | $21 \times 10^{-5}$ | $45 \times 10^{-6}$ | $36 \times 10^{-5}$ | $12 \times 10^{-6}$ | $18 \times 10^{-6}$ | $18 \times 10^{-6}$ | $75 \times 10^{-6}$ |
| diagonal | $33 \times 10^{-6}$ | $-65 \times 10^{-6}$ | $10 \times 10^{-5}$ | $18 \times 10^{-6}$ | $-19 \times 10^{-6}$ | $-19 \times 10^{-6}$ | $12 \times 10^{-6}$ |

**Table 8.** The comparative correlation coefficients of several ciphertext images of several methods.

| Ciphertext image | Direction | ReF. [12] | ReF. [13] | ReF. [15] | ReF. [18] | ReF. [21] | ReF. [23] | ReF. [24] | Proposed scheme |
|---|---|---|---|---|---|---|---|---|---|
| Lena | horizontal | $13\times10^{-3}$ | $10\times10^{-4}$ | $27\times10^{-4}$ | $14\times10^{-4}$ | $32\times10^{-4}$ | $22\times10^{-4}$ | $73\times10^{-5}$ | **$-29\times10^{-5}$** |
|  | vertical | $17\times10^{-3}$ | $-15\times10^{-4}$ | $13\times10^{-4}$ | $14\times10^{-4}$ | $16\times10^{-4}$ | $13\times10^{-4}$ | $44\times10^{-5}$ | **$21\times10^{-5}$** |
|  | diagonal | $67\times10^{-5}$ | $26\times10^{-4}$ | $11\times10^{-4}$ | $12\times10^{-4}$ | $23\times10^{-4}$ | $8\times10^{-4}$ | $36\times10^{-5}$ | **$33\times10^{-6}$** |
| Cameraman | horizontal | - | - | - | - | $14\times10^{-4}$ | $39\times10^{-4}$ | $93\times10^{-4}$ | **$27\times10^{-5}$** |
|  | vertical | - | - | - | - | $2\times10^{-4}$ | $7\times10^{-4}$ | $10\times10^{-5}$ | **$45\times10^{-6}$** |
|  | diagonal | - | - | - | - | $35\times10^{-4}$ | $86\times10^{-4}$ | $31\times10^{-4}$ | **$-65\times10^{-6}$** |
| Peppers | horizontal | - | $7\times10^{-5}$ | - | - | $6\times10^{-4}$ | $1\times10^{-4}$ | - | **$-80\times10^{-6}$** |
|  | vertical | - | $43\times10^{-4}$ | - | - | $38\times10^{-4}$ | $-26\times10^{-4}$ | - | **$12\times10^{-6}$** |
|  | diagonal | - | $-18\times10^{-4}$ | - | - | $10\times10^{-4}$ | $-23\times10^{-4}$ | - | **$18\times10^{-6}$** |
| Barbara |  | $-70\times10^{-4}$ | - | $-13\times10^{-4}$ | $12\times10^{-4}$ | - | - | - | **$86\times10^{-6}$** |
|  |  | $-79\times10^{-4}$ | - | $43\times10^{-4}$ | $27\times10^{-4}$ | - | - | - | **$18\times10^{-6}$** |
|  |  | $-22\times10^{-3}$ | - | $10\times10^{-4}$ | $-10\times10^{-4}$ | - | - | - | **$-19\times10^{-6}$** |

The correlation distribution of two pixels of Lena's plaintext and ciphertext images are illustrated in Fig. 9 for horizontal, vertical, and diagonal directions. Since the correlation distribution of a mono-color image would be a point and that of completely correlated pixels would be on $y = x$ line. As the correlation coefficients of Lena's plaintext image for the horizontal, vertical, and diagonal directions are 0.9737, 0.9838, and 0.9645, respectively; their correlation distributions are mainly concentrated on $y = x$ line. On the other hand, the correlation distributions of the ciphertext image are uniformly spread because of their very low correlation coefficients of $-29\times10^{-5}$, $21\times10^{-5}$, and $33\times10^{-6}$ (see Table 6).

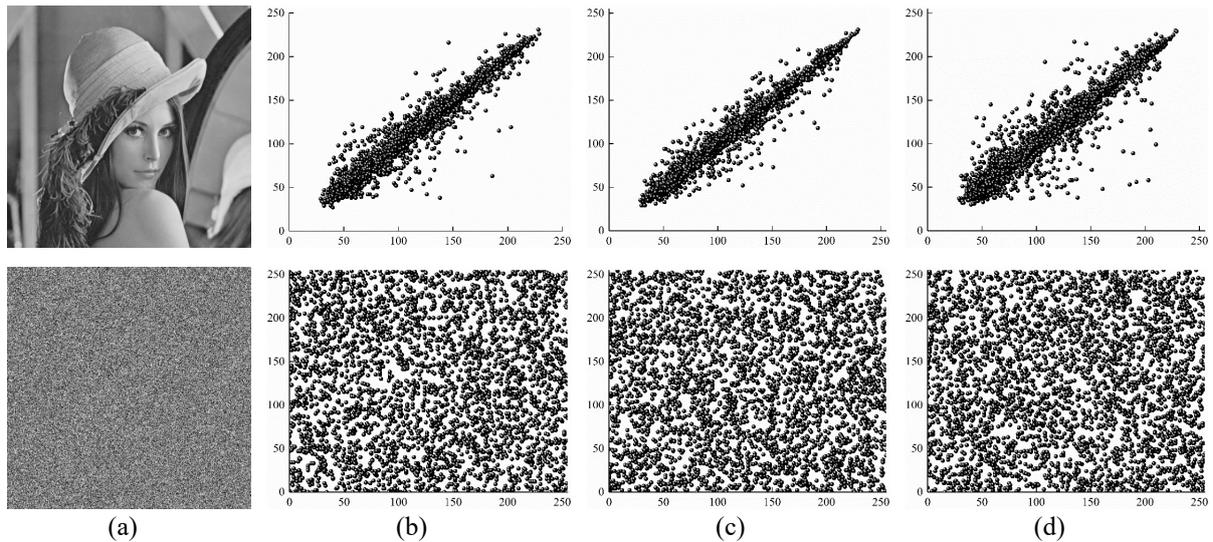

(a) (b) (c) (d)

**Fig. 9.** The correlation distribution of two adjacent pixels for three directions: a) the image, b) horizontal, c) vertical, d) diagonal

### 4.6 Differential attack analysis

A differential attack is attempted to learn the key and figure out the encryption scheme by tracing differences. Differential cryptanalysis examines the encryption scheme counter the cyberattacks through the difference between the plaintext and ciphertext images by altering a few bits in the plaintext image. It has capable of evaluating both permutation and manipulation performances of an encryption scheme. Therefore, a reliable encryption scheme should be sensitive to a slight change in the plaintext image. Differential attack analysis is carried out by calculating the following number of pixels changing rate (NPCR) and the unified average changing intensity (UACI).

$$D(i,j) = \begin{cases} 0, if\ C^1(i,j) = C^2(i,j) \\ 1, if\ C^1(i,j) \neq C^2(i,j) \end{cases} \quad (9)$$

$$NPCR = \frac{\sum_{i,j} D(i,j)}{mxn} x100\% \qquad (10)$$

$$UACI = \frac{1}{mxn}\left[\sum_{i,j} \frac{|C^1(i,j) - C^2(i,j)|}{255}\right] x100\% \qquad (11)$$

where $m$ and $n$ denote the height and width of the image. $C^1$ and $C^2$ are the ciphertext images before and after 1-bit of the plaintext image is altered, respectively. For a 1-bit altered grayscale image, the ideal scores of NPCR and UACI are expected to be 99.6094% and 33.4635%, respectively [41]. The scores of NPCR and UACI for under-analysis images encrypted through the proposed scheme are given in Table 9, and some of them are compared in Table 10 with the scores for Lena, Cameraman, Peppers, and Barbara available in the literature. It evident that the proposed scheme encrypts images with the closest results to the ideal scores.

**Table 9.** The scores of NPCR and UACI for under-analysis images encrypted through the proposed scheme.

| Direction | Lena | Cameraman | Baboon | Peppers | Barbara | Airplane |
|---|---|---|---|---|---|---|
| NPCR | 99.6094 | 99.6087 | 99.6094 | 99.6085 | 99.6085 | 99.6082 |
| UACI | 33.4622 | 33.4578 | 33.4652 | 33.4659 | 33.4631 | 33.4607 |

**Table 10.** The comparative scores of NPCR and UACI for several ciphertext images of several methods.

| Image | Test | ReF. [12] | ReF. [13] | ReF. [15] | ReF. [18] | ReF. [21] | ReF. [23] | ReF. [24] | Proposed scheme |
|---|---|---|---|---|---|---|---|---|---|
| Lena | NPCR | 99.5800 | 99.6000 | 99.6912 | 99.6621 | 99.6000 | 99.60934 | 99.6078 | 99.6094 |
|  | UACI | 33.4300 | 33.4700 | 33.5098 | 33.5278 | 33.5000 | 33.45969 | 33.4268 | 33.4622 |
| Cameraman | NPCR |  | - | - | - | 99.6000 | 99.60683 | 99.6323 | 99.6087 |
|  | UACI |  | - | - | - | 33.5500 | 33.44610 | 33.4096 | 33.4578 |
| Peppers | NPCR | - | 0.9960 |  |  | 99.6100 | 99.60576 | - | 99.6085 |
|  | UACI | - | 33.4600 |  |  | 33.5200 | 33.50204 | - | 33.4659 |
| Barbara | NPCR | 99.6100 | - | 99.6912 | 99.7501 | - | - | - | 99.6085 |
|  | UACI | 33.4300 | - | 33.5098 | 33.5102 | - | - | - | 33.4631 |

**4.7 Cropping attack analysis**

Some parts of the ciphertext images can be lost or abused by the cyberattacks during the network transferring. Cropping attack analysis can assess the competence of an encryption scheme regarding not only the permutation but also the manipulation performance. Hence a robust and stable encryption scheme can recover the cropped image with the minimum degeneration. In order to analyze the proposed encryption scheme, the encrypted Lena images cropped with ratios of 1/16, 1/4, and 1/2 are decrypted in Fig. 10, and the image cropped with 1/16 is also compared in Fig. 11 with various schemes suggested in the literature. From the visual results, the proposed encryption scheme effectively recovers the cropped images with the least deterioration even if it is cropped with 1/2 as well as it is corroborated by the comparison.

Furthermore, the Lena image is numerically appreciated the proposed scheme by calculating Peak Signal to Noise Ratio (PSNR) defined in Eq. (12) precisely measures image quality based on comparing to the uncropped plaintext images. Therefore, PSNR should be higher as much as possible for the lower degeneration. The results of PSNR for under-analysis images encrypted through the proposed scheme are listed in Table 11, and that for the Lena are compared with the existing results reported elsewhere in Table 12 [24,36]. Thanks to the high PSNR, the cropping performance of the proposed encryption scheme is even validated in addition to the visual results in Fig. 11.

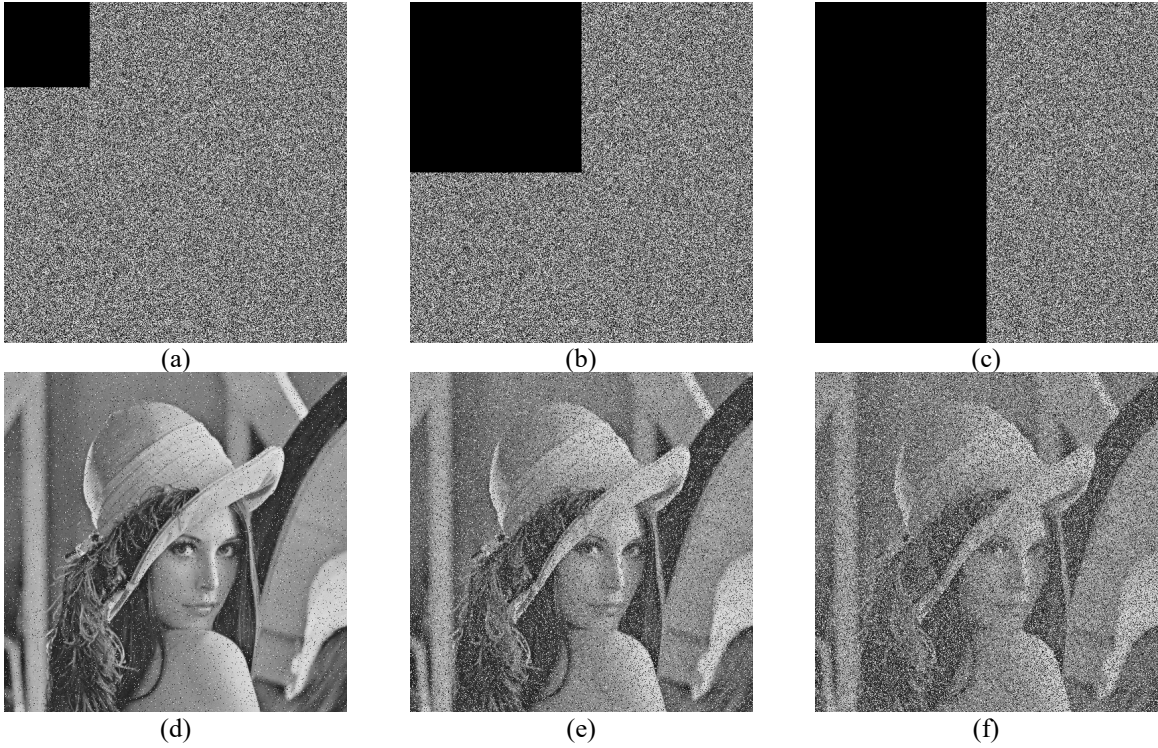

**Fig. 10.** Cropping attack analysis of the proposed encryption scheme using the Lena image, ciphertext images cropped with ratios a) 1/16, b) 1/4, c) 1/2, and decrypted images with ratios d) 1/16, e) 1/4, f) 1/2

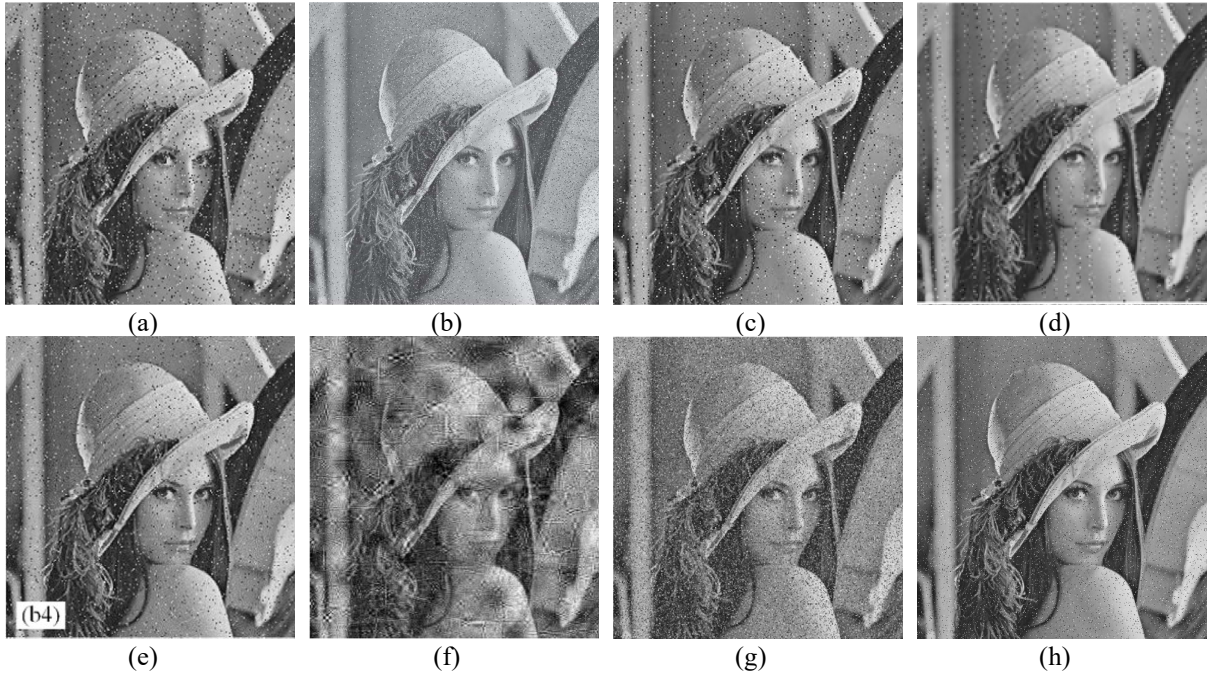

**Fig. 11.** Comparative cropping attack analysis of the encrypted Lena image cropped with 1/16 ratio (a) Ref. [23], (b) Ref. [24], (c) Ref. [34], (d) Ref. [35], (e) Ref. [36], (f) Ref. [37], (g) Ref. [38], (h) The proposed scheme

$$PSNR := 10\log\left(\frac{255^2}{MSE}\right) \quad (12)$$

where MSE is the mean squared error and defined as

$$MSE := \frac{1}{mn}\sum_{i=1}^{m}\sum_{j=1}^{n}(e_{ij} - f_{ij})^2 \quad (13)$$

where $E := [e_{ij}]$ is the plaintext image and $F := [f_{ij}]$ is the decrypted image that cropped.

**Table 11.** The results of PSNR metric for under-analysis images encrypted through the proposed scheme.

| Cropping ratio | Lena  | Cameraman | Baboon  | Pepper  | Barbara | Airplane |
|---|---|---|---|---|---|---|
| 1/16 | 21.37 | 20.5512 | 21.6142 | 20.9282 | 20.8795 | 20.1485 |
| 1/4  | 15.25 | 14.446  | 15.6453 | 14.8925 | 14.8560 | 14.0847 |
| 1/2  | 12.23 | 11.43   | 12.6529 | 11.8445 | 11.8296 | 11.0896 |

**Table 12.** The comparative results of the PSNR metric.

| Image | Cropping ratio | Ref [24] | Ref [36] | Proposed scheme |
|---|---|---|---|---|
|      | 1/16 | 17.58 | 20.78 | 21.37 |
| Lena | 1/4  | 15.03 | 14.96 | 15.25 |
|      | 1/2  | 12.13 | 12.08 | 12.23 |

### 4.8 Noise Attack Analysis

Some noise can be inherently added and/or inserted by the cyberattacks during the network transferring. Noise attack analysis is used to evaluate the permutation and manipulation performances of an image encryption scheme as similar to cropping attack analysis. Salt & pepper noise (SPN) is mostly utilized to inspect scheme against the noise attacks. In this way, restoring the ability of the scheme is investigated by adding SPN to the ciphertext image. Fig. 12 visually demonstrates the decrypted versions of the ciphertext images by adding various SPN densities of 0.001, 0.005, 0.01, and 0.1. Moreover, these results are numerically verified via PSNR metric measured as 40.23, 32.65, 29.23, and 19.23 for images with SPN densities of 0.001, 0.005, 0.01, and 0.1, respectively. Consequently, the proposed image encryption scheme maximally restores all images even if the image has a high SPN density such as 0.1.

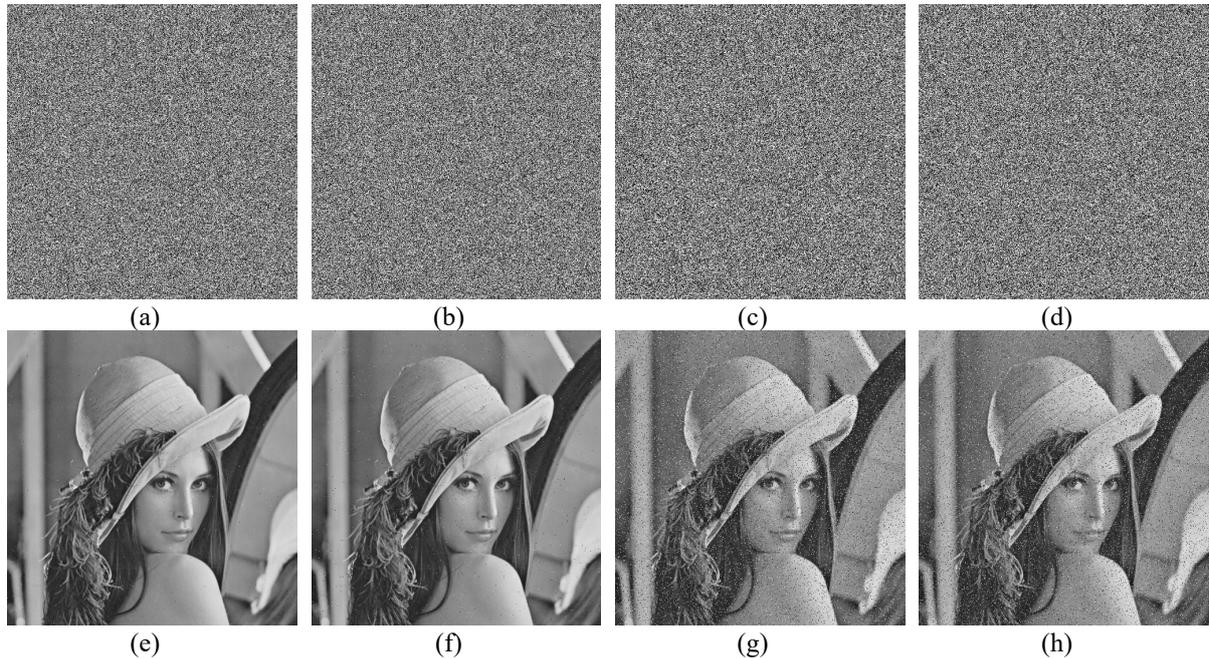

**Fig. 12.** Ciphertext images with adding various SPN densities of a) 0.001, b) 0.005, c) 0.01, d) 0.1 and their decrypted images with SPN densities of e) 0.001, f) 0.005, g) 0.01, h) 0.1

### 4.9 Computational Time Analysis

In addition to the cryptoanalysis, the computational time of an encryption scheme is an important aspect for a realistic image encryption scheme. Our scheme processes 0.015 s and 0.3996 s for deep CNN and encrypting operations, respectively. In other words, it encrypts throughout an image in 0.4146 s and decrypts in 0.3662 s. Therefore, it applies to real-time applications due to fast computational time.

## 5. Conclusion

In this study, an image encryption scheme is proposed, depending on a novel chaotic log-map, key generation via deep CNN, and a new bit reversion for encrypting operations. The images are securely encrypted across four operations: permutation, DNA encoding, diffusion, and bit reversion at which the pixels are scrambled and manipulated. The diverse hyperchaotic sequences for the operations are achieved by the log-map whose initial values and control parameters are produced by using the generated public key and secret key. The performance of the encryption scheme is visually and numerically investigated for the well-known images with respect to a variety of trusted crypto-analysis as well as validated by comparing with the available results in the literature. It is hence demonstrated that the proposed image encryption scheme is prominent among the suggested scheme due to the best absolute and unsurpassed results.